\documentclass[conference]{IEEEtran}
\IEEEoverridecommandlockouts
\usepackage[style=ieee]{biblatex}
\usepackage{amsmath,amssymb,amsfonts}
\usepackage{algorithmic}
\usepackage{graphicx}
\usepackage{textcomp}
\usepackage{xcolor}
\usepackage{multirow}
\usepackage{array}
\usepackage{caption}
\usepackage{subcaption}
\def\BibTeX{{\rm B\kern-.05em{\sc i\kern-.025em b}\kern-.08em
    T\kern-.1667em\lower.7ex\hbox{E}\kern-.125emX}}

\begin{document}

\title{Label-Efficient Sleep Staging Using Transformers Pre-trained with Position Prediction}

\author{\IEEEauthorblockN{Sayeri Lala}
\IEEEauthorblockA{\textit{Electrical and Computer Engineering} \\
\textit{Princeton University, Apple\textsuperscript{*}\thanks{*Work completed during internship at Apple.}}\\
Princeton, USA \\
slala@princeton.edu}
\and
\IEEEauthorblockN{Hanlin Goh}
\IEEEauthorblockA{\textit{Apple} \\
Cupertino, USA \\
hanlin@apple.com}
\and
\IEEEauthorblockN{Christopher Sandino}
\IEEEauthorblockA{\textit{Apple} \\
Cupertino, USA \\
csandino@apple.com}
}

\maketitle

\begin{abstract}
Sleep staging is a clinically important task for diagnosing various sleep disorders, but remains challenging to deploy at scale because it because it is both labor-intensive and time-consuming. Supervised deep learning-based approaches can automate sleep staging but at the expense of large labeled datasets, which can be unfeasible to procure for various settings, e.g., uncommon sleep disorders. While self-supervised learning (SSL) can mitigate this need, recent studies on SSL for sleep staging have shown performance gains saturate after training with labeled data from only tens of subjects, hence are unable to match peak performance attained with larger datasets. We hypothesize that the rapid saturation stems from applying a sub-optimal pretraining scheme that pretrains only a portion of the architecture, i.e., the feature encoder, but not the temporal encoder; therefore, we propose adopting an architecture that seamlessly couples the feature and temporal encoding and a suitable pretraining scheme that pretrains the entire model. On a sample sleep staging dataset, we find that the proposed scheme offers performance gains that do not saturate with amount of labeled training data (e.g., 3-5\% improvement in balanced sleep staging accuracy across low- to high-labeled data settings), reducing the amount of labeled training data needed for high performance (e.g., by 800 subjects). Based on our findings, we recommend adopting this SSL paradigm for subsequent work on SSL for sleep staging.

\end{abstract}


\section{Introduction}

Sleep staging is a clinically important task for diagnosing various sleep disorders affecting the population at large. Polysomnography (PSG) is the gold-standard approach for measuring sleep stages through overnight monitoring of the subject in the clinic using various sensing modalities including but not limited to electroencephalography (EEG). PSG data is manually scored by clinicians through a labor-intensive and time-consuming process \cite{berry2012aasm,malhotra2013performance}, making it difficult to scale this solution. Therefore, automatic labeling of PSG data is necessary for making it and other forms of data collection (e.g., using wearables) accessible. 

Recent studies have demonstrated that supervised deep learning-based solutions can automate sleep staging but require labeled training data collected from hundreds to thousands of subjects \cite{phan2022sleeptransformer,phan2022automatic}, which makes it untenable to deploy given the various types of target populations and data collection setups. While transfer learning can help \cite{phan2020towards}, this solution assumes access to a comparably large external labeled training dataset, which may not be available in practice nor similar to the target data distribution. 

Self-Supervised Learning (SSL) is a viable alternative that can reduce the need for large labeled training datasets by pretraining with self-generated tasks, like reconstructing masked inputs \cite{spathis2022breaking}. While few studies have shown that SSL can boost sleep staging performance over the baseline approach of training from randomly initialized weights (``scratch" training), these gains generally appeared only under settings with limited amounts of labeled data, i.e., fewer than tens of subjects \cite{banville2019self, eldele2023self,chien2022maeeg}; furthermore, these gains saturated and hence would be insufficient to match performance obtained under large labeled datasets \cite{banville2019self}. For example, the study in \cite{eldele2023self} pretrained a state-of-the-art (SOTA) sleep staging architecture, i.e., AttnSleep, \cite{eldele2021attention}, which comprises a CNN-based feature encoder and an attention-based temporal encoder \cite{vaswani2017attention}, and showed that it outperformed scratch training when the labeled training set came from less than tens of individuals but otherwise performed on par when fine-tuned with the complete labeled dataset, which only comprised 20 subjects. 

We think that this saturation of performance gains under SSL is explained by the use of a suboptimal pretraining scheme. In particular, the recent study examining SOTA architectures and SSL algorithms \cite{eldele2023self} pretrained only the feature encoder, following traditional pretraining schemes, e.g., SimCLR \cite{chen2020simple}, while training only the temporal encoder during fine-tuning. We hypothesize that pretraining both the feature and temporal encoders would boost performance across low- and high-labeled data regimes. To investigate this, inspired by the AttnSleep architecture, we adopt the Transformer \cite{vaswani2017attention}, which seamlessly integrates temporal and feature encoding through the use of repeated blocks, each comprising a multi-head attention layer followed by a fully-connected linear layer, and pretrain using a SOTA architecturally relevant SSL algorithm, i.e., Masked Patch Position Prediction (MP3) \cite{zhai2022position}. We compare this model against scratch training under low- and high-labeled data regimes to assess whether the performance gains persist throughout. On a sample sleep staging dataset, we find that the proposed pretraining scheme, which implicitly pretrains both the feature and temporal encoder by the coupled nature of the encoder, maintains performance gains under different training dataset sizes, which translate into large reductions in the amount of labeled training data needed (e.g., by 800 subjects). We conclude that pretraining the feature and temporal encoder helps preclude diminishing returns under SSL applied to sleep staging.

\section{Background}

The Transformer is a neural-network-based architecture that encodes sequential data; compared to Convolutional or Recurrent Neural Networks, Transformers can model long temporal relationships more effectively by avoiding the need to process the data sequentially, which contributes to vanishing gradients. To do this, the Transformer first tokenizes or reshapes the input sequence into a sequence of tokens or patches of equal size; then, it encodes the position of each patch by modulating the tokenized input using learned or hard-coded positional embeddings (e.g., sinusoid-based). This input is passed to an encoder with blocks comprising multi-head self-attention and fully-connected feed-forward layers. The multi-head self-attention transforms each patch by calculating its similarity to the other keys or patches and averaging the values or patches according to this similarity (i.e., dot-product-based attention); multiple heads are used to introduce variance in the attention activation, which are then aggregated. The embedding from the multi-head attention is normalized and then passed into a feed-forward block, which linearly transforms the embedding of each patch.  


MP3 \cite{zhai2022position} is a SOTA SSL algorithm in which a Transformer is trained to predict the position of each patch given a shuffled set of non-overlapping input patches extracted from an image with no positional encoding. As a result, the Transformer is required to learn both local and global relationships between patches to perform the position prediction task. The MP3 task is made more difficult by masking a subset of the input patches and using only unmasked keys and values for all patch queries. It was shown in \cite{zhai2022position} that additionally masking patches improved downstream performance compared to the scratch baseline on an image classification task; however, this strategy degraded performance on a speech classification task, highlighting inherent task difficulties in analyzing 1D vs 2D data under MP3.

\section{Methodology}

Here we adapt the MP3 pretraining algorithm to learn both local and global temporal features from sleep staging data. Given a shuffled set of non-overlapping  windows extracted from a 30s EEG signal, a Transformer model is trained to predict the temporal position of each window (i.e., a patch). We followed the strategy in \cite{zhai2022position} for applying MP3 to 1D signals; we did not mask windows and we provide some positional encoding to make it easier to learn from the MP3 task. We adopt hyperparameters for the Transformer, e.g., dimension of embedding layer, position encoding scheme, encoder depth, number of heads, dropout, etc., from the original architecture \cite{vaswani2017attention}, as described in Table \ref{tab:arch}. 


\begin{table}[htbp]
\caption{Transformer hyperparameters used in our study. Total number of parameters in the architecture is 18,986,661.}
\label{tab:arch}
\begin{center}
\begin{tabular}{|c|c|}
\hline
\textbf{Component} & \textbf{Value} \\
\hline
Input normalization scheme & Instance-wise \\ \hline 
Original input size & $\mathbb{R}^{3000}$ \\
\hline
Tokenized input size & $\mathbb{R}^{101\times30}$ \\ \hline 
Linear embedding dimension & 512 \\ \hline
Position encoding scheme & sinusoid \\ \hline
Encoder depth & 6 \\ \hline
Number of heads & 8  \\ \hline
Feedforward dimension & 2048 \\ \hline
Dropout rate & 0.1 \\ \hline 

\end{tabular}
\end{center}
\end{table}

\begin{figure}
     \centering
          \begin{subfigure}[b]{0.4\textwidth}
         \centering
         \includegraphics[width=\textwidth,scale=0.25,trim={2cm 0cm 2cm 0cm},clip]{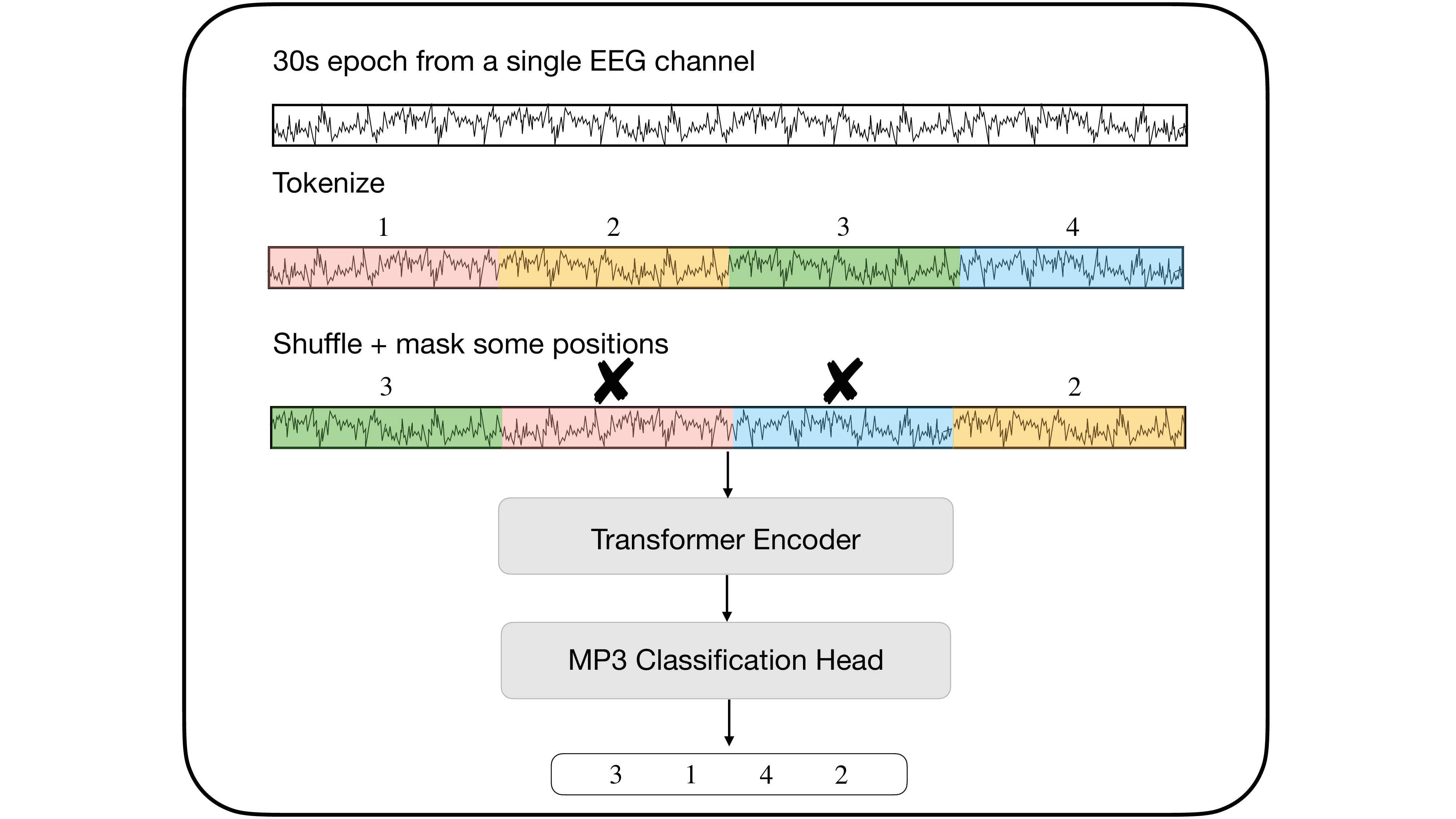}
         \caption{Scheme for pretraining the Transformer with MP3. In this example, the tokens are shuffled such that tokens 3, 1, 4, and 2 appear consecutively, and the original position is provided for tokens 3 and 2, while the position is masked out for tokens 1 and 4. MP3 trains the model to predict the original position per token from the shuffled input.}
         \label{fig:pretrain}
     \end{subfigure}
     
     \begin{subfigure}[b]{0.4\textwidth}
         \centering
         \includegraphics[scale=0.45,trim={2cm 0cm 2cm -3cm},clip,width=\textwidth]{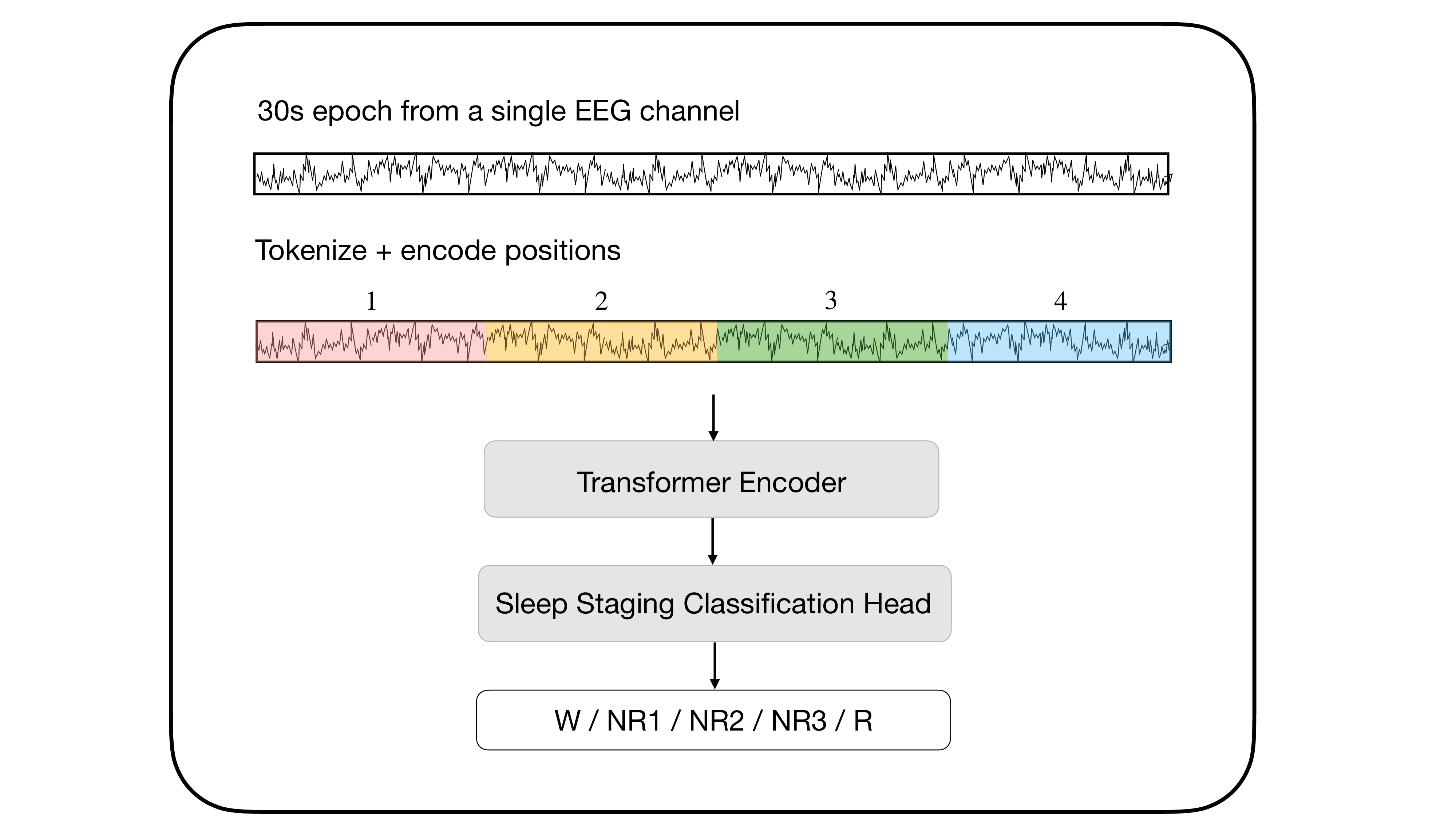}
         \caption{Scheme for supervised training of the Transformer. The tokens are ordered and all their positions are provided. The model is trained to predict the sleep stage for the EEG segment.}
         \label{fig:sup}
     \end{subfigure}
     \hfill
     \caption{Pretraining and supervised training schemes for our sleep staging Transformer model. For illustration purposes, the input is split into four tokens, indexed $1$-$4$.}

\end{figure}

To learn the model weights, we first pretrain the model using the MP3 objective with a pretext classification head and then fine-tune on the labeled dataset, replacing the pretext head with a classification head for sleep staging, with the schemes illustrated in Figures \ref{fig:pretrain} and \ref{fig:sup}, respectively. We conduct pretraining and fine-tuning using the Adam optimizer with a batch size of 512 for 50 and 200 epochs, respectively. During supervised training, we use the class-weighted cross-entropy loss, which weighs classes inversely proportional to their frequencies to mitigate class imbalance effects. For simplicity, we do not incorporate other regularization techniques, e.g., data augmentation and learning rate scheduling. We tune hyperparameters according to the best supervised validation performance. Specifically, for MP3, we set the learning rate to $10^{-3}$, masking ratio to 0, and amount of  positional information added to 50\%. For supervised training, we tune the learning rate over the interval between $10^{-5}$ and $10^{-3}$. We trained our models using four NVIDIA Tesla V100 GPUs with 32GB of memory.

\section{Experimental Setup}

We evaluate the proposed scheme by comparing its performance against the baseline scheme of scratch training. We use the PhysioNet 2018 Challenge dataset \cite{ghassemi2018you}, which contains PSG recordings from subjects monitored at Massachusetts General Hospital, where the EEG data were split into consecutive, non-overlapping 30s epochs, each labeled into one of five sleep stages, i.e., wake (W), rapid eye movement (REM), non-REM stage 1 (NR1), non-REM stage 2 (NR2), and non-REM stage 3 (NR3). Following similar setups under prior work \cite{banville2019self, eldele2023self}, we use the F3-M2 EEG channel downsampled from 200 to 100 Hz using Fourier re-sampling. We divide the dataset into training, validation, and test splits subject-wise, yielding splits with comparable class distributions, as shown in Table \ref{tab:data_split}. 

\begin{table}
\caption{Dataset splits for training and evaluation.}
\label{tab:data_split}
\begin{center}
\begin{tabular}{|c|c|c|c|}
\hline
\textbf{Split} & \textbf{\# Subjects} & \textbf{\# Instances} & \textbf{W/NR1/NR2/NR3/R} (\%) \\
\hline
Training & 657 & 591,473 & 18/15/42/12/13\\ \hline 
Validation	& 219	& 195,613 & 17/15/43/11/14 \\ \hline
Test 	&117	& 104,241 & 19/15/42/12/12 \\ \hline
\end{tabular}
\end{center}
\end{table}

Using the dataset, we conduct two experiments. In the first experiment, we use the same training dataset for pretraining and supervised training but vary the size of the training (and validation) datasets to compare performance differences between scratch training and pretraining followed by finetuning under low- and high-data regimes; specifically, we construct smaller training and validation datasets by taking one random sample with 1\% or 10\% of the number of subjects in the original training and validation datasets. In the second experiment, we increase the size of the pretraining dataset relative to that used for supervised training to characterize how performance varies as a function of the pretraining set size. 

To evaluate performance, we report standard classification metrics, e.g., balanced accuracy, overall accuracy, Cohen's Kappa, macro F1, and per-class F1, which are normalized to $[0,1]$ range.

\section{Results}

The results of our first experiment are reported in Table \ref{tab:exp_1}. When the size of the training and validation datasets are 1\% of the original splits, pretraining generally improves performance. Overall performance improves and classwise performance improves (e.g., on the W class) or remains competitive (e.g., NR1 and NR2 classes) to scratch. The performance gains under pretraining persist under higher data regimes as seen when the size of the training and validation datasets are increased to 10\% and 100\% of the original splits. The performance gains translate into large reductions in the amount of labeled training data required; for example, pretraining and fine-tuning with 10\% of the original training/validation splits reaches performance competitive to scratch training with 100\% of the training/validation splits.      

The performance gains under pretraining increase with the size of the pretraining dataset, as shown in the results of our second experiment, reported in Table \ref{tab:exp_2}. When only 1\% of the training/validation data are available during supervised training, increasing the pretraining data by 10-fold increases overall performance while generally maintaining or improving classwise performance. Increasing the pretraining data by 100-fold boosts performance further, making the model competitive to training from scratch with 10\% of the training/validation data. Performance gains were also observed when only 10\% of the training/validation data are available for supervised training; increasing the pretraining data by 10-fold makes the model outperform the model trained from scratch with 100\% of the training/validation data.






\begin{table*}[hbt!]
\caption{Experiment 1: Comparing the performance of the scratch and pretrained (PT) models under different amounts of training data. The pretrained model is trained on the data used for supervised training, without the labels. ``\% Supervised Training Data" indicates the amount of data used for training relative to the original split size.}
\label{tab:exp_1}
\begin{center}
\begin{tabular}{|c|c|c|c|c|c|c|c|c|c|c|}
\hline
\textbf{Method} & \textbf{\% Supervised Training Data} & \textbf{Bal. Acc.} & \textbf{Acc.} & \textbf{$\kappa$} & \textbf{MF1} & \textbf{W-F1} & \textbf{NR1-F1} & \textbf{NR2-F1} & \textbf{NR3-F1} & \textbf{R-F1} \\
\hline

\multirow{ 3}{*}{Scratch} & 1 & 0.47 & 0.44 & 0.29 & 0.43 & 0.49 & 0.36 & 0.47 & 0.62 & 0.20 \\
& 10 & 0.65&0.61&0.50&0.60&0.70&0.42&0.63&0.67&0.58 \\
& 100 & 0.71&0.68&0.59&0.67&0.74&0.46&0.73&0.72&0.71 \\ \hline

\multirow{ 3}{*}{PT} & 1 &0.52 & 0.47 & 0.33 & 0.47 & 0.59&0.35&0.47 &0.64 &0.30  \\
& 10 & 0.70&0.68&0.58&0.66&0.73&0.45&0.73&0.71&0.69 \\
& 100 & 0.74&0.71&0.62&0.70&0.76&0.50&0.75&0.74&0.76 \\ \hline

\end{tabular}
\end{center}
\end{table*}

\begin{table*}[hbt!]
\caption{Experiment 2: Comparing the performance of pretraining on data from the original labeled training dataset and pretraining with additional unlabeled data. The ``Amount of Additional Pretraining Data" is measured relative to the ``\% Supervised Training Data"; e.g., ``$1\times$" is a model pretrained on the same dataset used for supervised training and ``$10\times$" is a model pretrained with ten-fold more unlabeled data than labeled data.}
\label{tab:exp_2}
\centering
\begin{tabular}{|>{\centering\arraybackslash}m{1.5cm}|>{\centering\arraybackslash}m{1.5cm}|>{\centering\arraybackslash}m{1cm}|>{\centering\arraybackslash}m{1cm}|>{\centering\arraybackslash}m{1cm}|>{\centering\arraybackslash}m{1cm}|>{\centering\arraybackslash}m{1cm}|>{\centering\arraybackslash}m{1cm}|>{\centering\arraybackslash}m{1cm}|>{\centering\arraybackslash}m{1cm}|>{\centering\arraybackslash}m{1cm}|}

\hline
\textbf{\% Supervised Training Data} & \textbf{Amount of Additional Pretraining Data} & \textbf{Bal. Acc.} & \textbf{Acc.} & \textbf{$\kappa$} & \textbf{MF1} & \textbf{W-F1} & \textbf{NR1-F1} & \textbf{NR2-F1} & \textbf{NR3-F1} & \textbf{R-F1} \\
\hline

\multirow{ 2}{*}{1} & $1\times$ & 0.52 & 0.47 & 0.33 & 0.47 & 0.59&0.35&0.47 &0.64 &0.30 \\
& $10\times$ & 0.55&0.52&0.38&0.51&0.48&0.39&0.57&0.64&0.49  \\ 
& $100\times$ &0.63&0.60&0.48&0.60&0.63&0.44&0.65&0.69&0.57  \\ \hline

\multirow{ 2}{*}{10} & $1\times$ & 0.70&0.68&0.58&0.66&0.73&0.45&0.73&0.71&0.69 \\
& $10\times$ & 0.72&0.71&0.62&0.70&0.73&0.49&0.77&0.73&0.75 \\ \hline

\end{tabular}
\end{table*}


\section{Conclusion}

We have shown that pretraining an architecture that integrates feature and temporal encoding and fine-tuning it for sleep staging can boost performance across low- and high-data regimes, thereby reducing the amount of labeled data needed to match performance under large data regimes (i.e., labeled data from 1K subjects) by 90\% or approximately 800 subjects. Our findings suggest that this pretraining scheme overcomes the performance saturation observed under prior SSL sleep staging schemes by pretraining both the feature and temporal encoder; therefore, we propose that this scheme be a springboard for future work on SSL for sleep staging. Along these lines, it could be possible to improve performance further by adopting an architecture with a separate feature encoder followed by a separate temporal encoder (as in prior work \cite{eldele2021attention}) and developing a pretraining scheme that can jointly pretrain both encoders using encoder-specific objectives in order to enhance their synergy for the downstream task. 

\section*{Acknowledgments}

We are grateful to Shuangfei Zhai and Navdeep Jaitly for their guidance on the MP3 algorithm and to Ran Liu for helping with computational setup and discussions.

\printbibliography

\end{document}